\documentclass[12pt,a4paper]{article}

\newcommand{\email}[1]{{\small E-mail: #1}}

\newcommand{\Zo}{{\bf Z}}

\newcommand{\Tr}{\,{\rm Tr}\,}
\newenvironment{keywords}{\par\noindent Keywords: }{}
\newtheorem{condition}{Condition}

\begin{document}

\title{Continuity of a class of entropies and relative entropies}
\author{
Jan Naudts\\
\small Departement Natuurkunde, Universiteit Antwerpen,\\
\small Universiteitsplein 1, 2610 Antwerpen, Belgium\\
\email {Jan.Naudts@ua.ac.be}
}

\date{January 2004}

\maketitle

\begin{abstract}
The present paper studies
continuity of generalized entropy functions and relative entropies defined using the
notion of a deformed logarithmic function. In particular,
two distinct definitions of relative entropy are discussed.
As an application,
all considered entropies are shown to satisfy Lesche's stability condition.
The entropies of Tsallis' nonextensive thermostatistics are taken as examples.
\end{abstract}


\section{Introduction}

The discrete entropy functional
\begin{equation}
I_0(p)=-\sum_k p_k\log(p_k)\le\infty,
\label{shannon}
\end{equation}
is {\sl not} continuous in the total variation norm
\begin{equation}
||p-q||_1=\sum_k|p_k-q_k|,
\end{equation}
in case the number of {\sl microstates} $k$ is infinite. This means that a small
change in probability distribution may cause an arbitrary large
change in entropy. This discontinuity has been identified recently \cite {HT01}
as an essential characteristic of information content in natural languages.
But its occurrence can make it difficult to
obtain a reliable estimate of entropy from experimental observation.
In many cases the probabilities $p_k$ are defined
over a finite index set $k=1,2,\cdots,N$. Then uniform continuity
holds and a useful estimate, called Lesche's stability condition \cite {LB82},
exists --- see expression (\ref {lesche4}).
The inequality was already known before
since Fannes \cite {FM73} proved the quantum version of
the inequality about ten years earlier.
However, Lesche formulated the inequality as a condition which is
satisfied by (\ref {shannon}) but not by the alpha-entropies of R\'enyi \cite {RA65}.
Recently \cite {AS02}, it has been shown that also the
$q$-entropies of Tsallis' nonextensive thermostatistics
satisfy Lesche's condition. Here, we generalize this proof to a
large class of entropy functions, and formulate a more
general continuity estimate (\ref{cont1}).

It is known since long that in (\ref {shannon})
the natural logarithm may be replaced by an arbitrary
increasing function $f(x)$. The entropy of the 
discrete probability distribution function ({\bf pdf}) $p$
reads then
\begin{equation}
\tilde I(p)=-\sum_k p_kf(p_k).
\label{entdef}
\end{equation}
In the terminology of \cite {PD86,OP93,PD03} these are quasi-entropies.
It is clear that for general functions $f(x)$
not much can be said about continuity of entropy or relative entropy.
It is obvious to require that $f(x)$ shares some
of the properties of the natural logarithm. A class of functions
satisfying such extra conditions has been introduced recently \cite {NJ02}.
They have been used as the basis for a broad generalization
of thermostatistics \cite {NJ03, NJ04}. The present paper focuses on
entropy functionals occurring in this generalized thermostatistics.

A possible generalization of relative entropy, also called divergence \cite {KL51}, is
$f$-divergence \cite {CI63,CI72}, defined by
\begin{equation}
I(p||q)=\sum_kq_kf(p_k/q_k),
\label{relentdef}
\end{equation}
with $f(x)$ a convex function, defined for $x>0$, strictly convex at $x=1$.
The ratio $p_k/q_k$ can be seen as the discrete Radon-Nikodym
derivative of $p$ w.r.t.~$q$. The latter has been the basis for a systematic generalization
to the context of quantum mechanics --- see chapter 5 of \cite {OP93}.
Alternative expressions of the form
\begin{equation}
D(p||q)=\sum_k\big[f(p_k)-f(q_k)-(p_k-q_k)f'(q_k)\big],
\label{altdef}
\end{equation}
with $f'(x)$ the derivative of $f(x)$,
are called divergences of the Bregman type in the mathematics literature. In the
original definition \cite {BLM67} the pdfs $p$ and $q$ are interchanged.
Then (\ref {relentdef}) and (\ref {altdef}) are identical
in case $f(x)=x\log(x)$. Hence, in the standard theory there is no need to
make a difference between the two forms. To clarify why both are needed
let us remark that mean entropy, in contrast with dynamical entropy,
is negative relative entropy w.r.t.~some reference state. If the number $N$ of microstates is finite then
entropy is relative entropy w.r.t.~uniform probabilities $q_k=1/N$
\begin{equation}
-I(p||q)=-\frac{1}{N}\sum_kf(Np_k).
\label {entrelenet}
\end{equation}
The continuum limit of (\ref {entrelenet}) becomes
\begin{equation}
-I(p||q)\rightarrow -\int_0^1{\rm d}k\,f\big(\rho(k)\big)
\end{equation}
for any probability measure $p$ with density function $\rho(x)$
w.r.t.~the Lebesgue measure ${\rm d}x$ of $[0,1]$. This continuum
limit makes clear why a definition of relative entropy of the form
(\ref {relentdef}) is needed.
In what follows, the definition of generalized entropy that will be used is
\begin{equation}
I(p)=-\sum_kf(p_k).
\label {genentr}
\end{equation}
By omitting the factors $N$ from (\ref {entrelenet}) the explicit dependence
on the number of microstates disappears and the expression is of the form
(\ref {entdef}). In particular, if  $f(x)=x\log(x)$
then $I(p)$ coincides with $I_0(p)$.

There exist also situations where a divergence of the
form (\ref {altdef}) is needed. In (generalized) statistical
mechanics relative entropy $D(p||q)$ measures the difference in free
energy between an arbitrary pdf $p$ and the equilibrium pdf $q$.
The quantity $-f'(q_k)$ equals
the energy of the $k$-the microstate divided by temperature
(up to a constant term). Hence,
\begin{equation}
-\sum_kp_k\ln_\kappa(q_k)-I(p)
\end{equation}
is the (non-equilibrium) {\sl free energy} of $p$ divided by temperature $T$
(again up to a constant term). Then (\ref {altdef}) expresses that
free energy as a function of the pdf $p$ is minimal at equilibrium $p=q$.

In information theory the {\sl linking identity}
connects average code length, entropy and divergence
\begin{equation}
\langle\kappa,p\rangle=I(p)+D(p||q).
\label {li}
\end{equation}
See e.g.~\cite {HT01}. Here,
divergence measures the redundancy of the code $\kappa$ against the pdf $p$.
From (\ref {li}) follows
\begin{equation}
\langle\kappa,p\rangle-\langle\kappa,q\rangle
=I(p)+D(p||q)-I(q),
\end{equation}
which can be identified with (\ref {altdef}), provided that the
average code length is given by
\begin{equation}
\langle\kappa,p\rangle=-\sum_k p_kf'(q_k)+C,
\end{equation}
with $C$ a suitably chosen constant.

The paper is organized as follows.
The next section gives a short review of deformed exponentials and logarithms.
Sections 3, 4, and 5, discuss the definitions of entropy
and relative entropy.
Continuity estimates for entropy and relative entropy
are given in section 6. Finally, Lesche's stability condition
is discussed in sections 7 and 8. The paper is concluded with a short discussion
of results, followed by appendices, containing proofs of inequalities.

\section{Deformed exponentials and logarithms}

In \cite {NJ02}, a deformed logarithm is defined as
a strictly increasing concave function, defined
for all $x>0$, vanishing for $x=0$. Following \cite {NJ04}
it is written as
\begin{equation}
\ln_\phi(x)=\int_1^x{\rm d}y\,\frac 1{\phi(y)}
\end{equation}
with $\phi(y)$ a strictly positive increasing function. For convenience,
the integral of $\ln_\phi(x)$ is denoted
\begin{equation}
F_\phi(x)=\int_1^x\hbox{ d}y\,\ln_\phi(y)=\int_1^x{\rm d}y\frac {x-y}{\phi(y)}.
\label {Fdef}
\end{equation}
The possible divergence of $\ln_\phi(x)$
at $x=0$ should be mild enough so that $F_\phi(0)$ is finite.
The inverse function is the deformed exponential
$\exp_\phi(x)$ and is defined on the range of $\ln_\phi(x)$,
which may be less than the whole real line. If needed, the domain
of definition is extended by putting $\exp_\phi(x)=0$ if $x$
is too small, and $\exp_\phi(x)=+\infty$ if $x$ is too large.

For further use the notion of deduced logarithmic function
$\omega_\phi(x)$, associated with $\ln_\phi(x)$, is needed.
It is defined by
\begin{eqnarray}
\omega_\phi(x)
&=&(x-1)F_\phi(0)-x F_\phi(1/x)\cr
&=&x\int_0^{1/x}{\rm d}y\,\left(-\ln_\kappa(y)-F_\kappa(0)\right)\cr
&=&\int_0^{1/x}{\rm d}y\,\frac {xy-1}{\phi(y)}.
\end{eqnarray}
It is again a deformed logarithm provided that
\begin{equation}
\int_0^1\hbox{ d}x\,\ln_\phi(1/x)<+\infty.
\end{equation}

The name of $\kappa$-deformed logarithm is used in \cite {NJ02}
and, with a more restricted meaning, in \cite {KS01}.
To avoid confusion this name is used in the present paper
only with the latter restricted meaning.
Its origin is the kappa-distribu\-tion,
which is a generalization of the Maxwell distribution.
This distribution is given by
\begin{equation}
\rho(v)=A\left[1+\frac{1}{2\kappa}\beta\frac{v^2}{v_0^2}\right]^{-1-\kappa}
\end{equation}
and can be written as $\rho(v)=A\exp_{\phi}(-(1/2)\beta v^2/v_0^2)$
with the deformed logarithm $\ln_\phi(x)$ defined by
\begin{equation}
\ln_\phi(x)=\kappa(1-x^{-1/(1+\kappa)})
=\frac\kappa{1+\kappa}\int_1^x{\rm d}y\,y^{-(2+\kappa)/(1+\kappa)},
\qquad \kappa>0.
\end{equation}

As a simple example of deformed exponential
and logarithmic functions, consider the piecewise linear functions
determined by the values
\begin{equation}
\ln_\phi(a^n)=n,\qquad
\exp_\phi(n)=a^n,\qquad
n\in\Zo,
\end{equation}
with $a>0$ any base number.
But also the function $\ln_\phi(x)=-1+\sqrt x$ is a deformed
logarithm. Its inverse is given by $\exp_\phi(x)=0$ if $x\le -1$, and
$\exp_\phi(x)=(1+x)^2$ otherwise.

\section{Entropy}

The entropy $I_\phi(p)$ of a discrete pdf $p$ is
defined by means of the deduced logarithmic function
$\omega_\phi(x)$, rather than by the deformed logarithm
$\ln_\phi(x)$. The reason for doing so is that the derivative of
$\omega_\phi(x)$ exists and can be calculated in terms of $\ln_\phi(x)$
while not much is known in general about the derivative $1/\phi(x)$
of the function $\ln_\phi(x)$. The definition of entropy functional reads
\begin{equation}
I_\phi(p)=\sum_k p_k\omega_\phi(1/p_k)\le+\infty.
\label{ed}
\end{equation}
Note that the function $x\omega_\phi(1/x)$ is non-negative
and goes to zero in the limit $x=0$. Hence the expression
is well-defined.
Basic properties are $I_\phi(p)\ge 0$
and
\begin{equation}
I_\phi(\lambda p+(1-\lambda) q)
\ge\lambda I_\phi(p)+(1-\lambda)I_\phi(q),
\qquad 0\le\lambda\le 1,
\end{equation}
i.e.~entropy $I_\phi(p)$ is a concave function of the pdf $p$.

From the definition of the deduced logarithmic function $\omega_\phi(x)$
follows that
\begin{eqnarray}
I_\phi(p)
&=&\sum_k\left[(1-p_k)F_\phi(0)-F_\phi(p_k)\right]\cr
&=&-F_\phi(0)-\sum_k\int_0^{p_k}\hbox{ d}x\,\ln_\phi(x).
\label{ede}
\end{eqnarray}
In particular, $I_\phi(p)$ is of the form (\ref {genentr})
with
\begin{eqnarray}
f(x)=F_\phi(x)-(1-x)F_\phi(0).
\label {fFrel}
\end{eqnarray}

Let us discuss some examples.
If $\ln_\phi(x)$ is replaced by the natural logarithm $\log(x)$
then the entropy is denoted $I_0(p)$ and is given by the well-known
expression (\ref{shannon}).
As a further example, consider entropy
in the context of Tsallis' non-extensive thermodynamics
\cite {TC88}. Fix a number $\kappa$ between -1 and 1, not equal to 0.
A deformed logarithm is defined by
\begin{equation}
\ln_\phi(x)=(1+\kappa^{-1})(x^\kappa-1)
=\int_1^x{\rm d}y\,\frac{1+\kappa}{x^{1-\kappa}}.
\label{tsallislog}
\end{equation}
Note that this definition differs from the definition of
$q$-logarithm found in the Tsallis literature \cite {TC94},
which coincides with the deduced logarithm
\begin{equation}
\omega_\phi(x)=(1/\kappa)(1-x^{-\kappa}).
\label{omegatsal}
\end{equation}
A short calculation yields the entropy functional
\begin{equation}
I_\phi(p)=\frac{1}{\kappa}\left(1-\sum_kp_k^{1+\kappa}\right)
\end{equation}
This entropy functional was studied long ago by Havrda and Charvat \cite {HC67}
and by Dar\'oczy \cite {DZ70}. It is a monotonic function of R\'enyi's alpha-entropies \cite {RA65}.
It is the starting point of Tsallis' thermostatistics.
In the latter context it is common to use
the parameter $q=1+\kappa$ instead of $\kappa$.
In the present paper the symbols $p$, $q$, and $r$ are used for pdfs.

As a final example, consider the $\kappa$-deformed logarithm
introduced by Kaniadakis \cite {KS01, KG01}
\begin{equation}
\ln_\kappa(x)=\frac 1{2\kappa}(x^\kappa-x^{-\kappa}).
\end{equation}
The parameter $\kappa$ should satisfy $-1<\kappa<1$ to guarantee concavity
of the deformed logarithm. The inverse function reads
\begin{equation}
\exp_\kappa(x)=\left(\kappa x+\sqrt {1+\kappa^2 x^2}\right)^{1/\kappa}.
\end{equation}
The corresponding entropy functional is obtained directly from (\ref {ede}).
The result is
\begin{equation}
I_\kappa(p)=\frac 1{2\kappa(1+\kappa)}\left(1-\sum_kp_k^{1+\kappa}\right)
+\frac 1{2\kappa(1-\kappa)}\left(\sum_kp_k^{1-\kappa}-1\right).
\end{equation}

\section{Relative entropy}

Let $q$ be a pdf for which $q_k>0$ holds for all $k$
(this condition can be omitted if the deformed logarithm
is such that $\omega_\phi(0)$ is finite).
From (\ref{relentdef}) follows that the relative entropy of the pdf $p$,
given $q$, is defined by
\begin{equation}
I_\phi(p||q)=-\sum_kp_k\omega_\phi(q_k/p_k).
\label{red1}
\end{equation}
Note  that, using the definition of $\omega_\phi$,
one obtains
\begin{equation}
I_\phi(p||q)=\sum_k\int_{q_k}^{p_k}\hbox{ d}x\,\ln_\phi(x/q_k).
\label{rep}
\end{equation}
Expression (\ref {red1}) is of the form (\ref {relentdef}) with $f(x)$ given by
(\ref {fFrel}). In particular,
this means that the divergence $I_\phi(p||q)$, considered here,
is a special case of the $f$-divergence of \cite {CI63,CI72}, with
functions $f$ which are strictly convex and have a concave derivative.
Many properties of $f$-divergence are known --- see \cite {DS00}.
In particular, one has $I_\phi(p||q)\ge 0$
and $I_\phi(p||q)=0$ implies $p=q$. Also, $I_\phi(p||q)$ is jointly convex in $p$ and $q$.

For the example of Tsallis' entropy functional
one obtains, using (\ref{omegatsal}),
\begin{eqnarray}
I_\phi(p||q)
=\frac{1}{\kappa}\sum_kp_k\left(\left(\frac{p_k}{q_k}\right)^\kappa-1\right).
\label{tsallisrelent}
\end{eqnarray}
This expression has been introduced in the context of Tsallis' thermostatistics
independently by several authors \cite {AS98, TC98, SM98}. However, the
definition was known before in the context of R\'enyi's alpha-entropies --- see \cite {HH93}.

If $\ln_\phi(x)$ has a unique derivative $\ln'_\phi(x)=1/\phi(x)$ in the point $x=1$ and the probabilities $p_k$
depend on parameters $\theta^i$ then the generalized Fisher information metric \cite {AS85},
defined by $I_\phi(p+dp||p)=I_\phi(p||p+dp)=(1/2)g_{ij}(p){\rm d}\theta^i{\rm d}\theta^j$,
becomes
\begin{equation}
g_{ij}(p)=\ln'_\kappa(1)\sum_k p_k\frac{\partial \log(p_k)}{\partial \theta^i}
\frac{\partial \log(p_k)}{\partial \theta^j}.
\label {fish1}
\end{equation}
Note that this expression does not depend on the actual choice of deformed logarithm,
except through the prefactor $\ln'_\kappa(1)$.

\section{Alternative definition of divergence}

So far, definition (\ref{red1}) seems quite satisfactory. However,
as discussed in the introduction, there is a need for an alternative
definition of the form (\ref {altdef}). By modification of (\ref {rep}) one obtains
\begin{eqnarray}
D_\phi(p||q)
&=&\sum_k\int_{q_k}^{p_k}\hbox{ d}x\,
\left(\ln_\phi(x)-\ln_\phi(q_k)\right)\cr
&=&\sum_k\left[F_\phi(p_k)-F_\phi(q_k)-(p_k-q_k)\ln_\phi(q_k)\right]\cr
&=&I_\phi(q)-I_\phi(p)-\sum_k(p_k-q_k)\ln_\phi(q_k).
\label{red2}
\end{eqnarray}
This expression is of the form (\ref {altdef}) with $f(x)$ given by
(\ref {fFrel}).
Positivity of $D_\phi(p||q)$ follows immediately because $\ln_\phi(x)$
is an increasing function of $x$. Equality $D_\phi(p||q)=0$ implies
that $p=q$. Convexity in the first argument is straightforward.
For the example of Tsallis' entropy one obtains
\begin{equation}
D_\phi(p||q)=\frac{1}{\kappa}\sum_kp_k(p_k^\kappa-q_k^\kappa)
-\sum_k(p_k-q_k)q_k^\kappa,
\end{equation}
which is definitely different from (\ref {tsallisrelent}).

If the probabilities $p_k$
depend on parameters $\theta^i$ then the generalized Fisher information metric
becomes
\begin{equation}
g_{ij}(p)=\sum_k\ln'_\phi(p_k)\frac{\partial p_k}{\partial \theta^i}
\frac{\partial p_k}{\partial \theta^j}.
\label {fish2}
\end{equation}
Indeed, one has
\begin{eqnarray}
D_\phi(p+{\rm d}p||p)
&=&\sum_k\int_{p_k}^{p_k+{\rm d}p_k}\hbox{ d}x\,
\left(\ln_\phi(x)-\ln_\phi(p_k)\right)\cr
&=&\sum_k\int_{p_k}^{p_k+{\rm d}p_k}\hbox{ d}x\,
\left(\ln'_\phi(p_k)(x-p_k)+\cdots\right)\cr
&=&\frac 12\sum_k\ln'_\phi(p_k)\big({\rm d}p_k\big)^2+\cdots,
\end{eqnarray}
and similarly for $D_\phi(p||p+{\rm d}p)$.
In contrast with (\ref {fish1}) the metric tensor (\ref {fish2}) depends in a non-trivial
way on the deformed logarithm $\ln_\phi$.

\section{Continuity estimates of entropy and of relative entropy}

In Appendix A is proved that
\begin{eqnarray}
|I_\phi(p)-I_\phi(q)|
&\le&-\sum_k\int_0^{|p_k-q_k|}\ln_\phi(x)\hbox{ d}x\cr
&=&\sum_k\left[F_\phi(0)-F_\phi(|p_k-q_k|)\right]\cr
&\equiv&d(p,q)\le+\infty.
\label{cont1}
\end{eqnarray}
The r.h.s.~of (\ref{cont1}) defines a metric $d(p,q)$. In
particular, it satisfies the triangle inequality.
Note that the distance between two pdfs may be infinite.
This is not a problem since one can always define a new
metric by $d_M(x,y)=\min\{d(x,y),M\}$, with $M$ a fixed positive constant.
The two metrics $d$ and $d_M$ define the same topology.

If $\ln_\phi(x)$ is taken to be the natural logarithm $\ln(x)$ 
then (\ref{cont1}) becomes
\begin{equation}
|I_0(p)-I_0(q)|\le
||p-q||_1-\sum_k |p_k-q_k|\ln(|p_k-q_k|).
\end{equation}
More generally, take $\ln_\phi$ equal to the logarithm (\ref {tsallislog}), used in
the Tsallis context. Then (\ref{cont1}) becomes
\begin{equation}
|I_\phi(p)-I_\phi(q)|\le
(1+\kappa^{-1})||p-q||_1
-\kappa^{-1}\sum_k|p_k-q_k|^{1+\kappa}.
\label{lesche1}
\end{equation}

Differences in relative entropy can be estimated in a way
similar as for entropy differences. One finds (see Appendix A)
\begin{eqnarray}
\left|I_\phi(p||r)-I_\phi(q||r)\right|
&\le&d(p,q)+h_r(p,q)\cr
\left|D_\phi(p||r)-D_\phi(q||r)\right|
&\le&d(p,q)+e_r(p,q)
\label{relentcont}
\end{eqnarray}
with $d(p,q)$ as before, and with
\begin{eqnarray}
h_r(p,q)&=&\sum_k|p_k-q_k|\ln_\phi(1/r_k),\cr
e_r(p,q)&=&-\sum_k|p_k-q_k|\ln_\phi(r_k)
\label{hrdef}
\end{eqnarray}
The r.h.s.~of (\ref{relentcont}) is the sum of two distances,
each satisfying the triangle inequality.
Take $q=r$ in (\ref {relentcont}) to obtain an upper bound for $I_\phi(p||q)$,
resp.~$D_\phi(p||q)$.

\section{A general continuity condition}

The r.h.s.~of (\ref{cont1}) resembles the entropy of
a distribution with elements $|p_k-q_k|$. Introduce therefore
the symmetric difference $p\Delta q$ of two distinct pdfs
$p$ and $q$ by
\begin{equation}
(p\Delta q)_k=\frac{|p_k-q_k|}{||p-q||_1}.
\end{equation}
Note that $p\Delta q$ is again a pdf.
Its elements satisfy $(p\Delta q)_k\le 1/2$. This implies that
\begin{equation}
I_\phi(p\Delta q)\ge -F_\phi(0)-\ln_\phi(1/2).
\label {lb}
\end{equation}
In Appendix B is shown that from (\ref{cont1}) follows that
if $||p-q||_1\le 1$ then
\begin{eqnarray}
|I_\phi(p)-I_\phi(q)|
&\le&\frac{F_\phi(0)-F_\phi(||p-q||_1)}{F_\phi(0)}
\,\left[
F_\phi(0)+I_\phi(p\Delta q)
\right].
\label{improved}
\end{eqnarray}
If $||p-q||_1=1$, then this inequality coincides with (\ref{cont1}).

Take $\ln_\phi$ equal to the logarithm (\ref {tsallislog}), used in the
Tsallis context. Then (\ref{improved}) becomes
\begin{eqnarray}
|I_\phi(p)-I_\phi(q)|
&\le&\frac{1}{\kappa}\left[(1+\kappa)||p-q||_1
-||p-q||_1^{1+\kappa}
\right]
\left[1+I_\phi(p\Delta q)\right].\cr
& &
\end{eqnarray}
This is less sharp than (\ref {lesche1}) which can be written as
\begin{eqnarray}
|I_\phi(p)-I_\phi(q)|
&\le&\frac{1}{\kappa}(1+\kappa)||p-q||_1
+||p-q||_1^{1+\kappa}\left[
I_\phi(p\Delta q)-\frac{1}{\kappa}\right].\cr
& &
\end{eqnarray}
In combination with (\ref {lb}), (\ref {improved}) shows that the entropy functional
$I_\phi(p)$ satisfies the following condition.

\begin{condition}
For each $\epsilon>0$ there exists $\delta>0$ such that
\begin{eqnarray}
|I(p)-I(q)|
&\le&\epsilon\, I(p\Delta q)
\end{eqnarray}
holds for all pdfs $p$ and $q$ satisfying $p\not=q$ and $||p-q||_1\le\delta$.
\end{condition}

To show relevance of this condition one consequence is highlighted.
Note that $(\lambda p+(1-\lambda)q)\Delta q$ does not depend on
$\lambda$ in the range $0<\lambda\le 1$. Hence 
Condition 1 implies that for each $\epsilon>0$ there exists $\delta>0$
such that
\begin{eqnarray}
& &|I(\lambda p+(1-\lambda)q)-I(\mu p+(1-\mu)q)|
\le \epsilon\, I(p\Delta q)
\end{eqnarray}
holds for distinct pairs $p$ and $q$, and for all $\lambda$ and $\mu$ between 0 and 1,
satisfying $|\lambda-\mu|\,||p-q||_1\le\delta$.
This result implies uniform continuity of entropy on the
segment $(p,q)$, provided $I(p\Delta q)$ is finite.

\section{Lesche's stability condition}

Assume now that the number of microstates is finite, equal to $N$
(i.e., the index $k$ of the pdfs $p$ and $q$ runs from 1 to $N$).
Introduce the notation
\begin{equation}
I^{\rm max}(N)=\max\{I(p):\,p_k=0\hbox{ for }k>N\}.
\end{equation}
Lesche \cite {LB82} showed twenty years ago that $I_0(p)$
satisfies the following condition.
\begin{condition}
For each $\epsilon>0$ there exists $\delta>0$ such that
\begin{eqnarray}
|I(p)-I(q)|
&\le&\epsilon\, I^{\rm max}(N)
\label {c2}
\end{eqnarray}
holds for all pdfs $p$ and $q$ satisfying $||p-q||_1\le\delta$
and $p_k=q_k=0$ for $k>N$.
\end{condition}
It is clear that an entropy function $I(p)$ satisfying
Condition 1 also satisfies Condition 2. For fixed $N$ these conditions imply
uniform continuity, which is a rather trivial statement because a
continuous function on a compact set is automatically uniformly
continuous. In addition, (\ref {c2}) specifies how
the estimate depends on the number of nonzero components $N$.

In the remainder of this section some inequalities, used in the
literature to prove Lesche's condition, are shown to follow from (\ref {cont1}).
In Appendix C is shown that (\ref {cont1}) implies that
\begin{eqnarray}
|I_\phi(p)-I_\phi(q)|
&\le& NF_\phi(0)-NF_\phi(N^{-1}||p-q||_1)\cr
&=&-N\int_0^{||p-q||_1/N}{\rm d}x\ln_\phi(x)\cr
&=&||p-q||_1\left[F_\phi(0)+\omega_\phi(N/||p-q||_1)\right].
\label {cont2}
\end{eqnarray}
It is difficult to bound $\omega_\phi(N/||p-q||_1)$ by
$I_\phi^{\rm max}(N)=\omega_\phi(N)$ in the general case using only that $\omega_\phi(x)$
is a concave increasing function. However, in the case that the deformed
logarithm is given by (\ref {tsallislog}), then one has
\begin{eqnarray}
\omega_\phi(N/||p-q||_1)
&=&\frac{1}{\kappa}(1-||p-q||_1^{\kappa})+||p-q||_1^{\kappa}\omega_\phi(N).
\end{eqnarray}
This can be used to write (\ref {cont2}) in the following form
\begin{eqnarray}
|I_\phi(p)-I_\phi(q)|&\le&
(1+\kappa^{-1})||p-q||_1\cr
& &
+\left[-\kappa^{-1}+I_\phi^{\rm max}(N)\right]||p-q||_1^{1+\kappa}.
\label{lesche3}
\end{eqnarray}
This is the result obtained recently by Abe \cite {AS02}.
It implies that $I_\phi(p)$ satisfies Condition 2.
In the limit $\kappa=0$ (\ref{lesche3}) becomes
\begin{eqnarray}
|I_0(p)-I_0(q)|&\le&
\big(1+I_0^{\rm max}(N)\big)||p-q||_1
-||p-q||_1\ln(||p-q||_1).
\label{lesche4}
\end{eqnarray}
This is the expression obtained originally by Lesche \cite{LB82}.
Fannes \cite {FM73,FMcomment} showed that, if $||p-q||_1\le 1/3$, then one has
the slightly stronger inequality
\begin{eqnarray}
|I_0(p)-I_0(q)|&\le&
I_0^{\rm max}(N)||p-q||_1
-||p-q||_1\ln(||p-q||_1).
\label{fannes}
\end{eqnarray}

\section{Discussion}

The present paper considers a large class of entropy functionals.
Their definition is based on the concept of deformed
logarithms. These entropies have nice enough properties to
enable the proof of useful estimates.
Only discrete pdfs have been considered. Expressions for continuous
distributions and for quantum probabilities are found in \cite {NJp}.

For each entropy functional $I_\phi(p)$ there
exists a metric $d(p,q)$ majorizing the
difference $|I_\phi(p)-I_\phi(q)|$ --- see inequality (\ref{cont1}).
The difference of relative entropies
$|I_\phi(p||r)-I_\phi(q||r)|$ is majorized with the
sum of two distances, the distance $d(p,q)$ mentioned above, and a
distance $h_r(p,q)$ which depends on the pdf $r$
--- see (\ref{relentcont}, \ref{hrdef}).

An alternative definition of relative entropy $D_\phi(p||q)$
has been proposed. It satisfies similar properties as $I_\phi(p||q)$,
but serves other goals. It is used in generalized statistical physics to measure
changes in free energy. In information theory it is a measure of redundancy.

Although the proof of (\ref{cont1}) is rather elementary, the
result can be used to show that all entropy functionals, considered
in the present paper, satisfy Lesche's stability condition (Condition 2 of the paper),
as well as a stronger version of the inequality (Condition 1 of the paper).
The proof is shorter and more transparent than that of \cite {AS02}.

\section*{Acknowledgement}
I thank Dr. P. Harremo\"es and Prof. H. Hasegawa
for providing some of the references to the literature.

{
  \appendix
  \renewcommand{\theequation}{A\arabic{equation}}
  \setcounter{equation}{0}  
  \section*{Appendix A}

Here we prove the inequalities (\ref{cont1}) and (\ref{relentcont}).
Consider
\begin{equation}
I_\phi(p)-I_\phi(q)
=-\sum_k\int_{q_k}^{p_k}\hbox{ d}x\,\ln_\phi(x).
\label{entdif}
\end{equation}
If $p_k<q_k$ then the contribution is negative and may be omitted
when trying to obtain an upperbound. Hence
one gets immediately, using Heavisides function $\theta(x)$,
\begin{eqnarray}
I_\phi(p)-I_\phi(q)
&\le&-\sum_k\theta(p_k-q_k)\int_{0}^{p_k-q_k}\hbox{ d}x\,\ln_\phi(x)\cr
&\le&-\sum_k\int_{0}^{|p_k-q_k|}\hbox{ d}x\,\ln_\phi(x).
\end{eqnarray}
This proves (\ref{cont1}).

To prove (\ref{relentcont}) note that from (\ref {rep}) follows
\begin{eqnarray}
I_\phi(p||r)-I_\phi(q||r)
&=&\sum_k\int_{q_k}^{p_k}\hbox{ d}x\,\ln_\phi(x/r_k).
\label{relentdif}
\end{eqnarray}
Assume $p_k<q_k$  and write the $k$-th term as
\begin{equation}
-\int_{p_k}^{q_k}\hbox{ d}x\,\ln_\phi(x/r_k).
\end{equation}
It increases when $-\ln_\phi(x/r_k)$ is replaced by $-\ln_\phi(x)$.
Hence the sum of all these terms is less than $d(p,q)$. 
On the other hand, if $p_k\ge q_k$ then the factor
$\ln_\phi(x/r_k)$ in the $k$-th term can be replaced
by $\ln_\phi(1/r_k)$, which yields the bound
\begin{equation}
\int_{q_k}^{p_k}\hbox{ d}x\,\ln_\phi(x/r_k)
\le(p_k-q_k)\ln_\phi(1/r_k).
\end{equation}
The sum of these terms is bounded by $h_r(p,q)$.
This finishes the proof of (\ref{relentcont}a).

In case of the alternative definition of divergence one has
\begin{equation}
D_\phi(p||r)-D_\phi(q||r)
=-I_\phi(p)+I_\phi(q)-\sum_k(p_k-q_k)\ln_\phi(r_k).
\end{equation}
Hence, in this case the estimate is straightforward.

}

{
  \appendix
  \renewcommand{\theequation}{B\arabic{equation}}
  \setcounter{equation}{0}  
  \section*{Appendix B}

Here, expression (\ref{improved}) is derived.
Note that any increasing
concave function $g(x)$, satisfying $g(0)\ge 0$,
also satisfies
\begin{equation}
g(\lambda x)g(y)\le g(x)g(\lambda y)
\end{equation}
for all $\lambda$, $x$, and $y$, for which $0\le \lambda\le 1$
and $0<x<y$ hold. Apply this result with $g(x)=F_\phi(0)-F_\phi(x)$
(which is increasing on $0\le x\le 1$),
$\lambda=||p-q||_1$, $x=|p_k-q_k|/||p-q||_1$, and $y=1$.
Note that the assumption $||p-q||_1\le 1$ is needed here.
There follows, using $F_\phi(1)=0$,
\begin{eqnarray}
& &\hskip -1cm\left[F_\phi(0)-F_\phi(|p_k-q_k|)\right]F_\phi(0)\cr
&\le&\left[F_\phi(0)-F_\phi(|p_k-q_k|/||p-q||_1)\right]
\left[F_\phi(0)-F_\phi(||p-q||_1)\right].
\end{eqnarray}
Using (\ref{cont1}) this implies (\ref{improved}).

}

{
  \appendix
  \renewcommand{\theequation}{C\arabic{equation}}
  \setcounter{equation}{0}  
  \section*{Appendix C}

Here, inequality (\ref {cont2}) is proved.
Because $F_\phi(x)$ is convex one has for any $x$ and $a>0$
\begin{equation}
F_\phi(x)\ge F_\phi(a)+(x-a)\ln_\phi(a).
\end{equation}
Therefore (\ref {cont1}) implies
\begin{equation}
|I_\phi(p)-I_\phi(q)|
\le NF_\phi(0)-NF_\phi(a)-\left(||p-q||_1-Na\right)\ln_\phi(a).
\end{equation}
The optimal choice of $a$ is $a=N^{-1}||p-q||_1$. This implies (\ref {cont2}).

}



\begin{thebibliography}{99}
\bibitem {FMcomment} The quantum mechanical entropy of a state with
density matrix $\rho$ is defined by $I_0(\rho)=-\Tr\rho\log\rho$.
The inequality
\begin{eqnarray}
|I_0(\rho)-I_0(\sigma)|\le \lambda\log N-\lambda\log\lambda
\nonumber
\end{eqnarray}
holds with $\lambda=\Tr|\rho-\sigma|$ whenever $\lambda\le 1/3$.
Here, $N$ is the dimension of the Hilbert space, Tr is the trace.
The inequality is quoted in \cite {OP93}, together with a proof,
in Proposition 1.8, be it with quite different notations.
\end{thebibliography}
\end{document}